%% file: paper.tex
\documentclass[sigconf, screen, nonacm]{acmart}
\AtBeginDocument{%
  \providecommand\BibTeX{{%
    \normalfont B\kern-0.5em{\scshape i\kern-0.25em b}\kern-0.8em\TeX}}}

\setcopyright{rightsretained}
\copyrightyear{2025}
\acmYear{2025}
\acmDOI{10.1145/1122445.1122456}

\usepackage{soul}
\input{macro}
\begin{document}

\title[\systemName: User-Assisted 3D Gaussian Splatting Avatar Refinement with Automatic Pose Suggestion]{\systemName: User-Assisted 3D Gaussian Splatting Avatar Refinement with Automatic Pose Suggestion}

\author{Jotaro Sakamiya}
\orcid{0009-0000-2010-1472}
\affiliation{%
 \institution{The University of Tokyo}
 \country{Japan}
}
\author{I-Chao Shen}
\orcid{0000-0003-4201-3793}
\affiliation{%
 \institution{The University of Tokyo}
 \country{Japan}
}
\author{Jinsong Zhang}
\orcid{0000-0001-9619-5030}
\affiliation{%
 \institution{Tianjin University}
 \country{China}
}
\author{Mustafa Doga Dogan}
\orcid{0000-0003-3983-1955}
\affiliation{%
 \institution{Adobe Research}
 \country{Switzerland}
}
\author{Takeo Igarashi}
\orcid{0000-0002-5495-6441}
\affiliation{%
 \institution{The University of Tokyo}
 \country{Japan}
}
\input{01_abstract}



\begin{teaserfigure}
  \includegraphics[width=\textwidth]{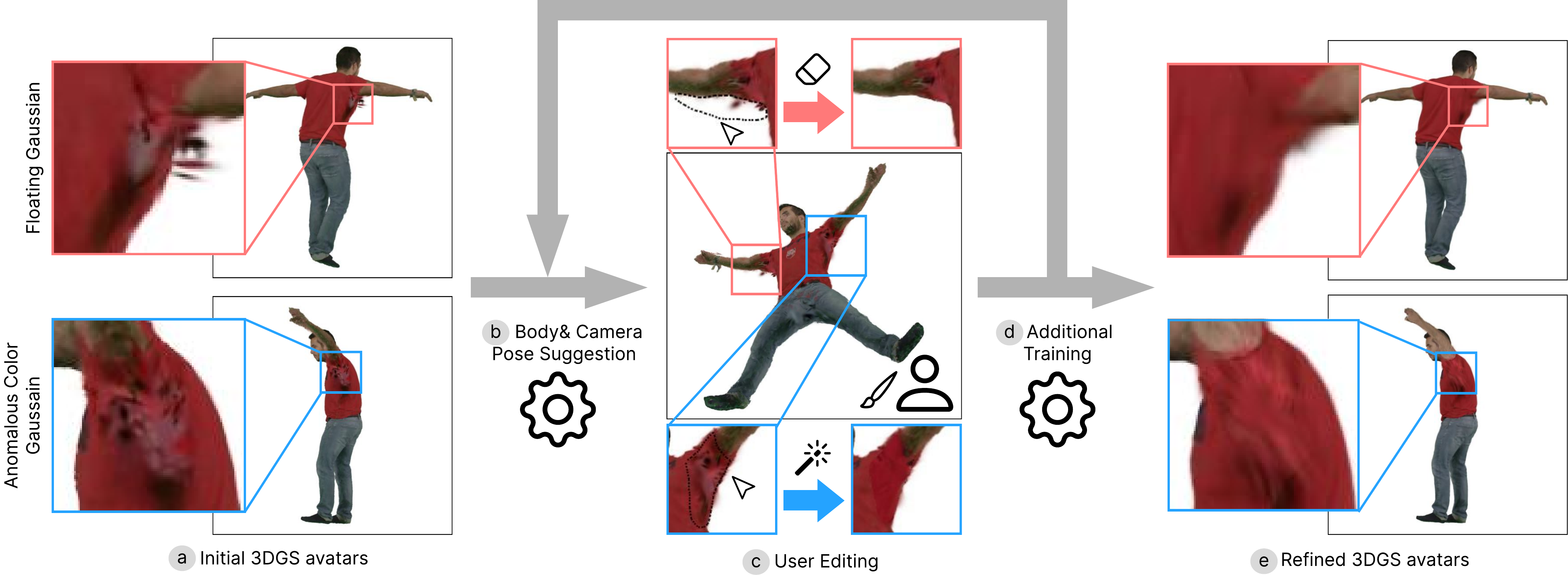}
  \caption{
    We present \systemName, a system for refining artifacts in 3D Gaussian Splatting (3DGS) avatars.
    Given (a) an initial 3DGS avatar with artifacts, (b) our system first renders a new image with the suggested body and camera poses. (c) Users can correct the artifacts on the rendered 2D image using the tools provided by our system. 
    (d) The refined image is used to update the avatar. After a couple of refinement iterations, (e) the refined 3DGS avatar has fewer artifacts and is more accurate under novel poses.
  }
  \label{fig:teaser}
\end{teaserfigure}

\maketitle

\input{02_introduction}
\input{03_relatedwork}

\input{04_prelim}
\input{05_overview}

\input{06_method}
\input{07_result}
\input{08_study}

\input{09_discussion}
\input{10_conclusion}

\bibliographystyle{ACM-Reference-Format}
\bibliography{paper}

\end{document}

%% file: macro.tex
\usepackage{enumitem}
\usepackage{amsmath}
\usepackage{amsfonts}
\usepackage{multirow}
\usepackage{bigdelim}
\usepackage{color}
\usepackage{booktabs}
\usepackage{ifthen}
\usepackage{hyperref}
\usepackage{subcaption}
\usepackage{bbold}
\usepackage{gensymb}
\usepackage[noabbrev,capitalise,nameinlink]{cleveref}
\creflabelformat{equation}{#2\textup{#1}#3}%
\usepackage{xcolor, pgf}
\usepackage{color, colortbl}
\usepackage[normalem]{ulem} 
\usepackage{placeins}
\usepackage{soul}
\usepackage[skins]{tcolorbox}
\usepackage{adjustbox}
\usepackage{multicol}

\usepackage[noabbrev,capitalise,nameinlink]{cleveref}
\creflabelformat{equation}{#2\textup{#1}#3}%

\definecolor{gray}{rgb}{0.5,0.5,0.5}
\definecolor{green}{rgb}{0, 0.6, 0}
\definecolor{orange}{rgb}{1, 0.5, 0}
\definecolor{mahogany}{rgb}{0.75, 0.25, 0.0}
\definecolor{purple}{rgb}{0.6, 0, 0.6}
\definecolor{darkgreen}{rgb}{0, 0.3, 0}
\definecolor{orange}{rgb}{1, 0.5, 0.}
\definecolor{lightblue}{rgb}{0.52, 0.75,0.91}
\definecolor{secondblue}{rgb}{0.52, 0.78,0.70}

\definecolor{softgreen}{rgb}{0.66,0.87,0.74}
\definecolor{softred}{rgb}{0.96,0.71,0.69}

\newcommand{\bestcell}[1]{\cellcolor{lightblue!50}#1}

\colorlet{soullightblue}{lightblue!50}
\newcommand{\besthint}[1]{\sethlcolor{soullightblue}\hl{#1}}
\colorlet{soulsecondblue}{secondblue!50}

\newcommand{\ie}{i.e.,}
\newcommand{\eg}{e.g.,}
\DeclareMathOperator*{\argmin}{arg\,min}

\usepackage{gensymb} 
\usepackage{xspace} 
\newcommand{\systemName}{\textsc{AvatarPerfect}\xspace}

\graphicspath{ {figures/} }

%% file: 01_abstract.tex
\begin{abstract}
Creating high-quality 3D avatars using 3D Gaussian Splatting (3DGS) from a monocular video benefits virtual reality and telecommunication applications.
However, existing automatic methods exhibit artifacts under novel poses due to limited information in the input video.
We propose \systemName, a novel system that allows users to iteratively refine 3DGS avatars by manually editing the rendered avatar images.
In each iteration, our system suggests a new body and camera pose to help users identify and correct artifacts.
The edited images are then used to update the current avatar, and our system suggests the next body and camera pose for further refinement.
To investigate the effectiveness of~\systemName, we conducted a user study comparing our method to an existing 3DGS editor \textit{SuperSplat}~\cite{supersplat}, which allows direct manipulation of Gaussians without automatic pose suggestions.
The results indicate that our system enables users to obtain higher quality refined 3DGS avatars than the existing 3DGS editor.
\end{abstract}

%% file: 02_introduction.tex
\section{Introduction}
\label{sec:intro}



Realistic and animatable 3D human avatars play an important role in many applications, such as AR/VR telepresence and games.
The quality of these avatars has been significantly improved due to recent advancements in novel radiance field representations such as NeRF~\cite{mildenhall2020nerf} and 3D Gaussian Splatting (3DGS)~\cite{kerbl3Dgaussians}.
With 3DGS, users can create lifelike 3DGS avatars by capturing an RGB monocular video covering a single body pose~\cite{moreau2024human,lei2024gart,hu2024expressive}. 
The resulting avatars can then be animated and rendered in various poses in real time.
The unprecedented quality and ease of use have made the 3DGS avatar appealing and accessible even for non-professional users. 

\begin{figure*}[ht!]
    \centering
    \includegraphics[width=\linewidth]{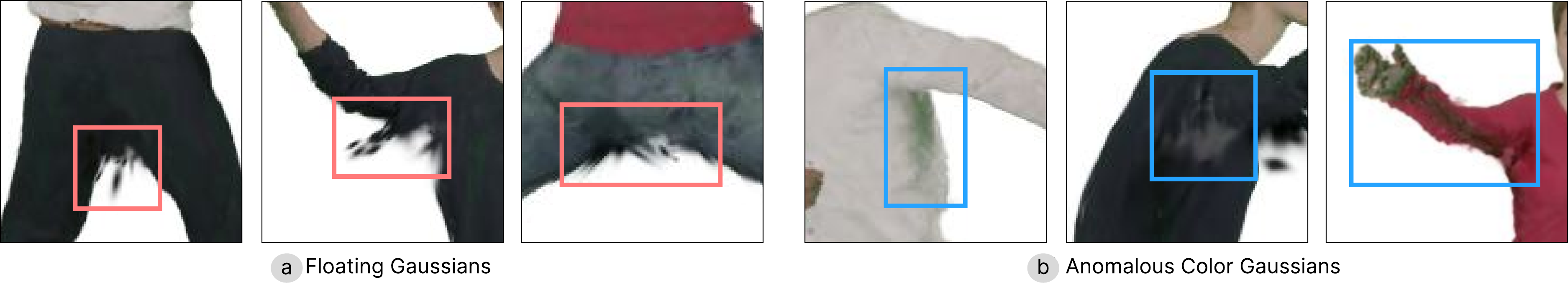}
    \caption{
    \textbf{Common artifacts in 3DGS avatars trained from a monocular input video.} 
    (a) Floating Gaussians: These arise in the occluded parts of the input video, where accurate estimation of Gaussian skinning weights is not feasible.
    (b) Anomalous Color Gaussians: These predominantly occur in the occluded parts of the input video, leading to unnatural Gaussian colors.
    }
    \label{fig:artifact}
\end{figure*}

However, 3DGS avatars generated by existing methods using a monocular video fail to properly capture and reconstruct the captured person under novel body poses.
Specifically, when we assign a novel body pose that is not captured in the input video to a 3DGS avatar, we often encounter two types of artifacts as shown in~\cref{fig:artifact}.
The first type of artifact is \textit{floating Gaussians}, where unnatural ellipses float around the avatar's body (red box in~\cref{fig:artifact}a).
This artifact occurs because it is challenging to obtain appropriate skinning weights for deforming the Gaussians using only a limited number of poses in the input video.
The second type of artifact is \textit{anomalous color Gaussians}, which creates shadow appearances or inconsistent colors in areas due to insufficient samples in the input monocular video (blue box in~\cref{fig:artifact}b).
This artifact occurs because the input videos often contain occluded parts of the subjects, making it challenging to accurately estimate the color of Gaussians in those hidden areas.
Consequently, when we assign novel poses to a generated 3DGS avatar, these artifacts appear around the hidden areas in the input video.
Despite previous algorithms~\cite{lei2024gart,hu2024gaussianavatar,shao2024splattingavatar} have attempted to eliminate these artifacts, they still persist when novel poses are applied, as these algorithms lack the reference information of novel poses.

In contrast, when presented with a static avatar under a novel pose not included in the input video, humans can still quickly identify these artifacts and manually manipulate artifact Gaussians using a simple 3D selection interaction~\cite{supersplat}.
However, there are two main issues with this interaction.
First, it is not intuitive for inexperienced users to manipulate 3D objects.
Second, due to the complex rendering process of 3DGS, it is unclear what the consequences of manipulating Gaussians are.
Specifically, directly manipulating the Gaussians in 3D space can lead to a less smooth appearance around the edited regions.
Therefore, it is challenging for users to determine the appropriate viewpoint for editing and which Gaussian to edit to reduce the artifacts.

To address this issue, we propose~\systemName, an interactive system for iteratively refining 3DGS avatars through 2D image editing.
Our system takes a 3DGS avatar, pre-trained with a monocular video and containing artifacts, as input.
During each iteration, \systemName suggests a new body and camera pose pair and renders an avatar image with artifacts for the user to refine (\cref{fig:teaser}c).
The user can remove the artifacts presented in the avatar image using various 2D image editing tools: background tool, inpaint tool, and diffusion-inpaint tool.
After the user removes artifacts in the suggested avatar images, \systemName updates the 3DGS avatar using both the original input video and the images edited by the user.
After several iterations, the resulting 3DGS avatar can be animated and rendered with reduced artifacts under novel poses (\cref{fig:teaser}e).

We formulate our body and camera pose suggestion method as a next best view (NBV) estimation problem~\cite{connolly1985determination} and built on the observation that the artifacts are more likely to occur in the hidden areas in the input video.
Therefore, we first measure the visibility of each Gaussian in the input video and avatar images refined by the user.
With visibility measurements, our method optimizes the next body and camera pose to display as many Gaussians as possible that were barely visible in the previously observed body and camera poses.
Our pose suggestion method alleviates the user's need to identify poses with artifacts and eliminates the need for complicated 3D manipulations.

We conducted a user study to compare the effectiveness of \systemName with an existing 3DGS editor: \textit{SuperSplat}~\cite{supersplat}.
In addition, we conducted a crowdsourcing study to assess the quality of the refined avatars generated by \systemName and \textit{SuperSplat}.
The results indicate that \systemName enables users to refine more artifacts and produce 3DGS avatars with higher quality.
We believe that our method, which allows users to effectively refine pre-trained 3DGS avatars, can reduce the effort required to generate high-quality 3D avatars and enhance VR and telecommunication experiences.


%% file: 03_relatedwork.tex
\section{Related Work}
\label{sec:related}
\subsection{Gaussian Splatting Avatars}
Recently, 3D Gaussian Splatting (3DGS)~\cite{kerbl3Dgaussians} has made significant progress in modeling and rendering novel views for both static scenes~\cite{yu2024mip,charatan2024pixelsplat} and dynamic scenes~\cite{wu20244d,li2024spacetime}.
With its powerful capacity and efficient rendering speed, 
recent methods have adopted Gaussian Splatting as the representation for modeling human avatars from a monocular video.
Some works~\cite{moreau2024human,lei2024gart,hu2024expressive} directly employ Gaussian Splatting as the canonical representation and then use a learnable linear blend skinning module to achieve deformation.
Another line of work~\cite{li2024animatable,pang2024ash,zheng2024gps} adopts the 2D UV map as the pose representation to predict the properties of 3D Gaussians using a 2D neural network. 
While these approaches have demonstrated promising results for observed poses and viewpoints in the input video, they still face challenges in generating realistic images for novel poses and hidden regions.

In this paper, we propose an interactive system to assist users in refining the 3DGS avatar.
Unlike previous automatic methods that focus on learning artifacts' visual features through additional complex learning models, our system allows users to directly identify and refine artifacts.


\subsection{3D Gaussian Splatting Editing}
SuperSplat~\cite{supersplat} enables the users to directly edit the Gaussians included in objects represented by 3D Gaussian Splatting. 
However, SuperSplat only allows for the direct removal of Gausians, making it unsuitable for more complex edits like changing the Gaussian colors.
Additionally, due to the complex rendering system of 3DGS, direct removal of Gaussians can result in an unexpected appearance.
Some works~\cite{zhuang2024tip, wu2024gaussctrl, wang2024gaussianeditor, chen2024gaussianeditor} edit the 3D scenes represented by 3DGS with text-prompts. 
In addition, other works~\cite{zhuang2024tip, chen2024gaussianeditor} enable users to edit the 3DGS scenes with image prompts.
Users enter a prompt and can control the 3D scene with style transformations and object additions.
However, the existing methods to edit the 3D scenes with text prompts and image prompts are for making the 3D scenes more controllable and not for repairing the artifacts in the 3D scenes.

\subsection{Next Best View Estimation}
Estimating the next best view (NBV) has been studied since~\cite{connolly1985determination} and is still actively studied in computer graphics, computer vision, and robotics.
The typical NBV method involves measuring task-specific uncertainty and proposing a view that minimizes this uncertainty the most.
Existing NBV estimation methods are used for active 3D geometry reconstruction methods~\cite{chen2024gennbv,qual_poisson,dhami2023pred, mendoza2020supervised}, 3D object classification~\cite{hou2024learning,wu20153d}, 3D texture painting~\cite{nbv_painting, chen2023text2tex} and robot manipulation~\cite{naazare2022online, jin2023neu, bircher2016receding}. 
Our pose suggestion method not only suggests camera pose but also body pose by measuring the visibility of Gaussians as our task-specific uncertainty, thereby reducing the user's effort in refining the 3DGS avatar.

\subsection{User-assisted Machine Learning}
With significant advancements in machine learning methods, these approaches are increasingly being applied in various real-world scenarios. 
However, automatic machine learning models often produce unsatisfactory results due to limited observations or vast solution spaces. 
Consequently, many user-assisted methods have been proposed to support existing machine learning models or data-driven algorithms across different applications.
For example, image classification performance can be enhanced by manipulating attention maps in deep neural networks~\cite{guide_atten} or providing task-specific labels~\cite{takahama2018adaflock}.
Similarly, object segmentation can be improved through the interactive provision of segmentation masks~\cite{zhou2022gesture}, and human pose estimation from video can be enhanced through 2D user manipulation~\cite{liu2024ipose}.
On the other hand, other works modeled user preference selections to optimize the parameters for visual content design~\cite{koyama2017sequential,koyama2022bo}, lighting design~\cite{yamamoto2022photographic}, melody composition~\cite{zhou2021interactive}, color map design~\cite{hong2024cieran}, and the quality of responses generated by large language models~\cite{brack2023illume}.
In contrast to previous methods, \systemName explores a new domain, specifically refining pretrained 3DGS avatars.
Our system helps users efficiently identify artifacts in 3DGS avatars and allows them to interactively refine them by directly editing the rendered 2D avatar images.

%% file: 04_prelim.tex
\section{Preliminary of 3D Gaussian Avatar}
\label{sec:prelim}
\begin{figure}
    \centering\includegraphics[width=\linewidth]{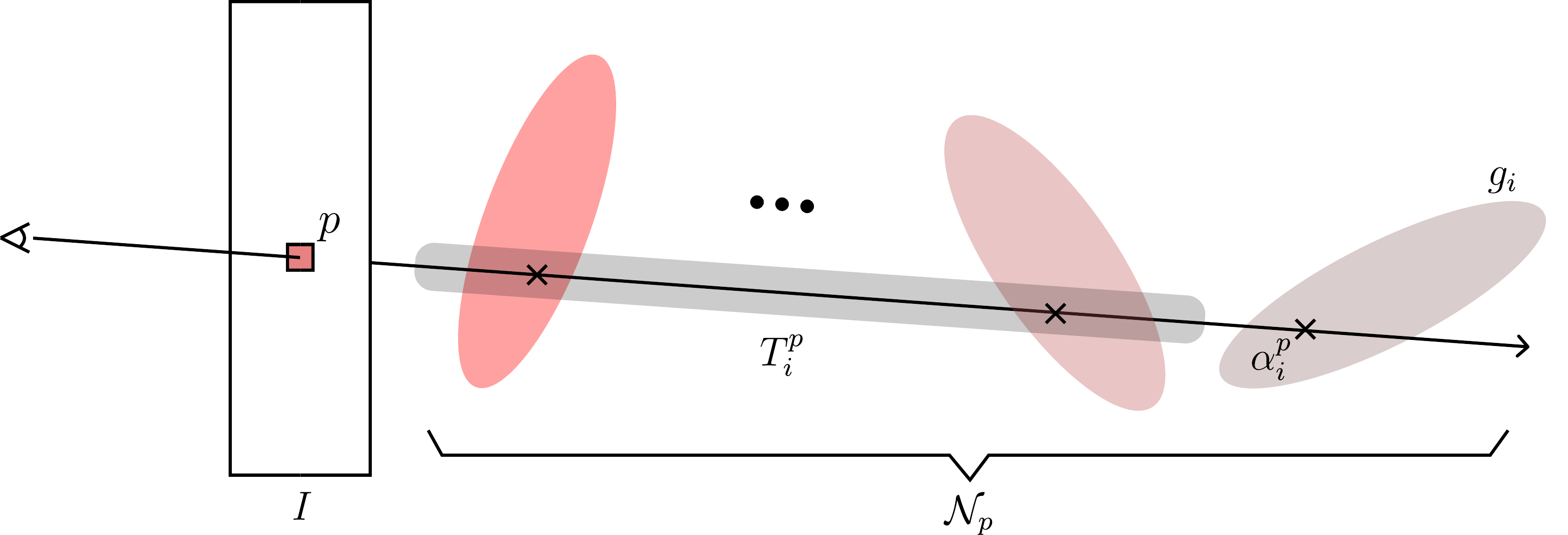}
    \caption{
    \textbf{Concept of 3DGS Rendering.} 
    In 3D Gaussian Splatting, we represent 3D scenes and objects as a collection of Gaussians. 
    When rendering a 2D image $I$, we determine the color of each pixel $p$ with a set of ordered Gaussians $\mathcal{N}_p$ that intersect with the ray corresponding to pixel $p$. 
    First, we calculate the contribution of each Gaussian $g_i$ to pixel $p$ as the product of $T_i$, the transmittance of the Gaussians located in front of $g_i$, and $\alpha_i^p$, the $\alpha$-blending value derived from the value of Gaussian $g_i$ at the point where it intersects the ray.
    We then determine the color of pixel $p$ by applying $\alpha$-blending based on the contribution of each Gaussian $g_i$.
    }
    \label{fig:gaussian_visibility}
\end{figure}

\begin{figure*}
    \centering\includegraphics[width=\linewidth]{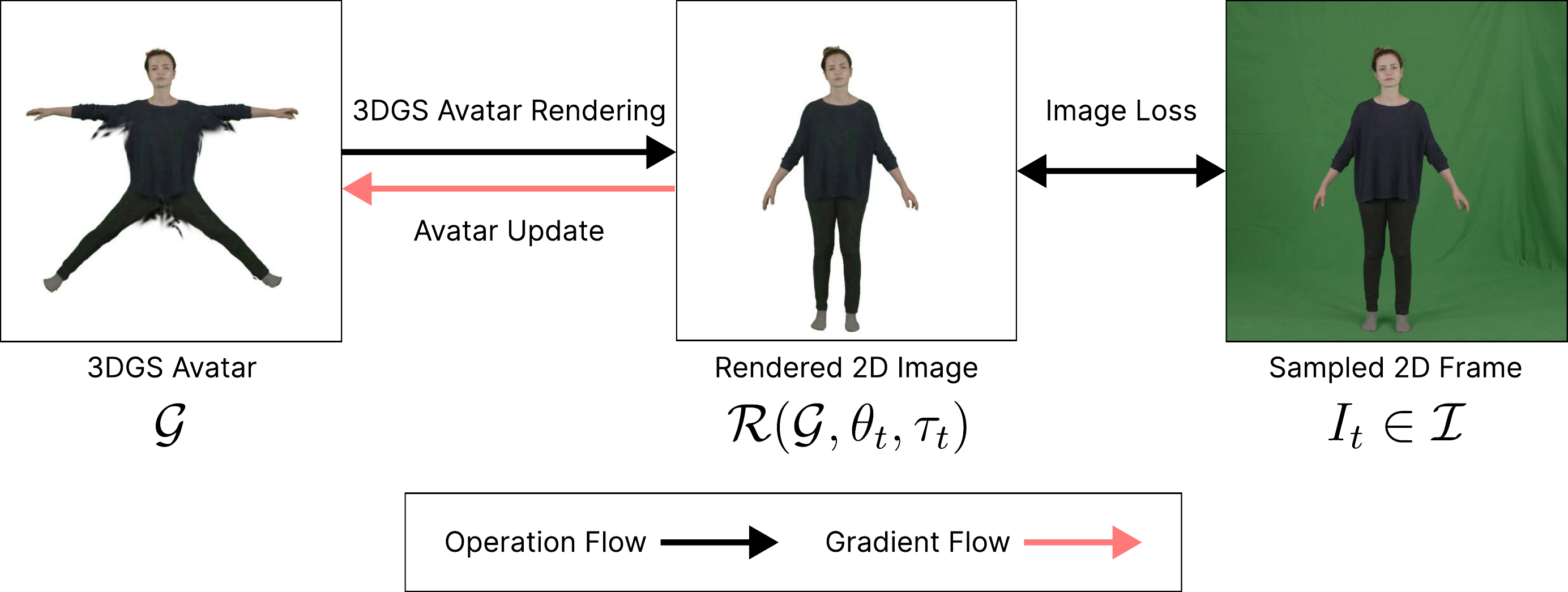}
    \caption{
    \textbf{General 3DGS avatar training pipeline.}
    In training a 3DGS avatar, the process iteratively repeats the calculation of image loss $L_\text{img}$ between the rendered image and the sampled image, followed by backpropagation of the gradients, similar to the original 3DGS.
    First, we sample a frame $I_t$ from the input video $\mathcal{I}$.
    At this point, each frame holds the body pose parameter $\theta_t$ of the subject and the camera pose parameter $\tau_t$.
    Using these parameters $\theta_t$ and $\tau_t$, we render the 3DGS avatar $\mathcal{G}$ into a 2D image by the rendering function $\mathcal{R}$.
    We then calculate the image loss $L_\text{img}$ between the rendered 2D image $\mathcal{R}(\mathcal{G}, \theta_t, \tau_t)$ and the sampled image $I_t$, and backpropagate the gradients to optimize the 3DGS avatar itself.
    In addition, we prune and densify the Gaussians in the 3DGS avatar according to the gradient information, as in the original 3DGS.
    }
    \label{fig:gaussian_avatar_training}
\end{figure*}
Before introducing the interface and next best pose suggestion method used in~\systemName, we briefly introduce the overview of 3DGS avatar. 
Given an input video $\mathcal{I}$, the goal of fitting a 3DGS avatar is to obtain an avatar that faithfully reconstructs the person's movement in the video.
In this paper, we represent the $t$-th video frame by a triad: a 2D image $I_t$, the avatar's body pose $\theta_t$ represented by SMPL~\cite{SMPL:2015}, and the camera pose $\tau_t$.
According to this notation, we denote the input video consisting of $T$ frames as $\mathcal{I} = \{(I_1, \theta_1, \tau_1), (I_2, \theta_2, \tau_2), \dots, (I_T, \theta_T, \tau_T)\}$.

Specifically, the 3DGS avatar is represented as a set of 3D Gaussians $\mathcal{G}=\{g_1, g_2, ..., g_n\}$, which can be rendered in real-time via differentiable rasterization.
Each 3D Gaussian $g_i$ $=(\mu_i, \sum_i, \eta_i, \mathbf{f}_i)$ is defined by its center location $\mu_i$, covariance matrix $\sum_i$, opacity $\eta_i$, and the spherical harmonics coefficients $\mathbf{f}_i$ for representing the view-dependent color.
Given the body pose $\theta_t$ and the camera pose $\tau_t$, we first deform the Gaussians $\mathcal{G}$ according to the body pose $\theta_t$ using Linear Blend Skinning (LBS)~\cite{LBS} and project them to a 2D plane using the camera pose $\tau_t$.
Then, the pixel color $C_p$ is computed by blending Gaussians that overlap at the given pixel $p$, sorted according to their depth (\cref{fig:gaussian_visibility}):
\begin{gather}
C_p = \sum_{i \in \mathcal{N}_p} T_i^p \alpha_i^p c_i^p, \label{eq:original_color} \\
T_i^p = \prod_{j = 1}^{i - 1}(1 - \alpha_j^p)\quad\text{and}\quad\alpha_i^p = \eta_i \exp{\biggl( -\frac{1}{2}(x_i^p - \mu_i)^T \Sigma_i^{-1} (x_i^p - \mu_i) \biggl)}
\end{gather}
where $\mathcal{N}_p$ is a set of ordered Gaussians that overlaps the pixel $p$, $c_i^p$ is the color of the Gaussian $g_i$ at pixel $p$ computed from the spherical harmonics coefficients $\mathbf{f}_i$, and $x_i^p$ is the intersection position between the camera ray of pixel $p$ and the Gaussian $g_i$.

Finally, as illustrated in~\cref{fig:gaussian_avatar_training}, we optimize the 3DGS avatar for each 2D frame $(I_t, \theta_t, \tau_t)$ from the input video $\mathcal{I}$.
At each optimization iteration, we randomly sample a 2D frame $(I_t, \theta_t, \tau_t)$ from the input video and compute the following loss function:
\begin{align}
    L_{\text{img}}(\mathcal{R}(\mathcal{G}, \theta_t, \tau_t), I_t), 
\label{eq:gs_fit}
\end{align}
where $\mathcal{R}$ is the rendering function that computes the pixel color using~\cref{eq:original_color}, and $L_{\text{img}}$ is the image loss that includes a series of image losses, including $L1$ distance and 1 - SSIM.
Additionally, during the 3DGS avatar update, we control the density of Gaussians in the 3DGS avatar by pruning and densification, similar to the optimization of the original 3DGS.



%% file: 05_overview.tex
\section{System Overview}
\label{sec:system_ui}

\begin{figure}
    \centering
    \includegraphics[width=\linewidth]{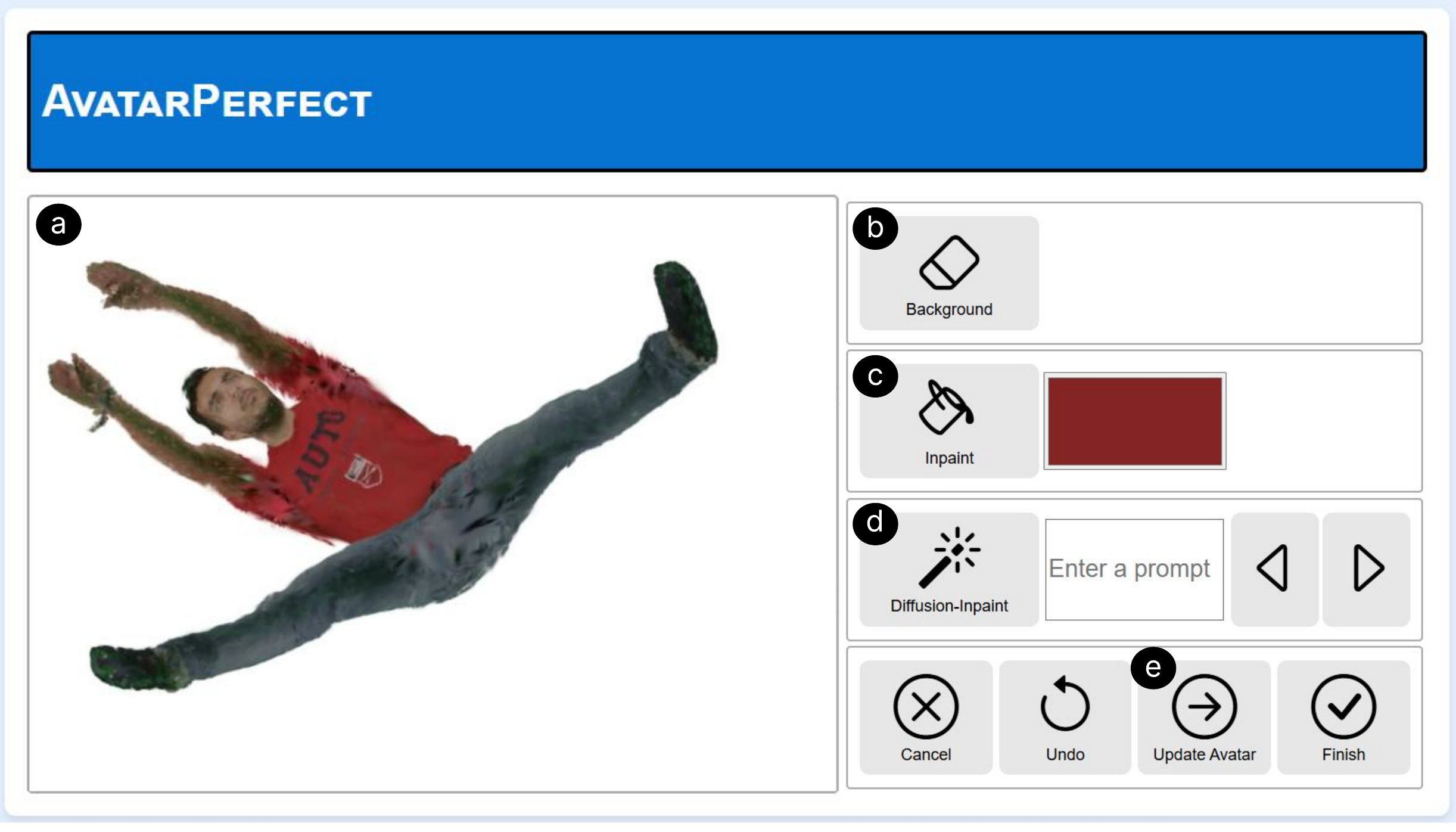}
    \caption{
    \textbf{User interface of \systemName.}
    In (a), users view the image rendered with the suggested body and camera pose.
    Our system provides three tools to edit the suggested image: (b) background tool, (c) inpaint tool, and (d) diffusion-inpaint Tool. 
    The background tool enables users to erase the floating Gaussians by painting selected regions white, which our system internally treats as background. 
    With the inpaint tool, users can paint the selected regions with a chosen color to address the anomalous color Gaussians. 
    The diffusion-inpaint Tool serves functionality similar to the inpaint tool but utilizes a diffusion model to paint the selected region.
    After their refinements on the suggested image, they can click (e) ``Update Avatar'' button to prompt our system to update the avatar with the edited image. After the 3DGS avatar update, our system suggests another avatar image for further refinement.
    }
    \label{fig:user_interface_edit}
\end{figure}


We propose a system, \systemName, for refining 3DGS avatars through iterative 2D image editing by the user. 
We provide a 2D image editing interface because editing a 3DGS avatar in 3D space is challenging for two reasons.
First, selecting the Gaussians in 3D space is not straightforward due to depth consideration. 
Second, the consequences of editing a Gaussian are often unclear due to its complex rendering process described in~\cref{sec:prelim}.
Therefore, our interface focuses on enabling users to refine the 3DGS avatar through 2D image editing.

The input to \systemName is a pretrained 3DGS avatar with artifacts under novel poses, not included in the input video.
Users can use our editing interface (\cref{fig:user_interface_edit}) to refine the 3DGS avatar until the artifacts are removed.
Our editing interface mainly consists of an image editing area (\cref{fig:user_interface_edit}a), three editing tools  (\cref{fig:user_interface_edit}b,c,d), and a button for the avatar update (\cref{fig:user_interface_edit}e).
Once the user inputs the 3DGS avatar, \systemName starts to show a rendered image of the current 3DGS avatar in the image editing area.
Then, the user can identify any artifacts and select an editing tool from the editing tool area to apply edits to the rendered image.
After making edits, the user can click the ``Update Avatar'' button to request \systemName to update the avatar with the edited image. After the 3DGS avatar update, our system suggests another avatar image in the image editing area to make further edits.

%% file: 06_method.tex
\section{System Features}
In this section, we introduce the details of (a) the 2D image editing tools that \systemName offers for the users to refine the avatar images, (b) how \systemName suggests avatar images to refine for users, and (c) how the avatar is updated after avatar image refinement by the user.

\subsection{2D Image Editing Tools}
\label{sec:user_painting}
When a user is provided with an avatar image rendered by~\systemName using a suggested body and camera pose, they can use 2D image editing tools to paint over any artifacts and produce a refined version of the image.
Then, \systemName uses the refined image and the painted mask to update the 3DGS avatar.
In~\systemName, we provide three types of tools: the background tool, the inpaint tool, and the diffusion-inpaint tool.

\begin{figure*}
    \centering
    \includegraphics[width=\linewidth]{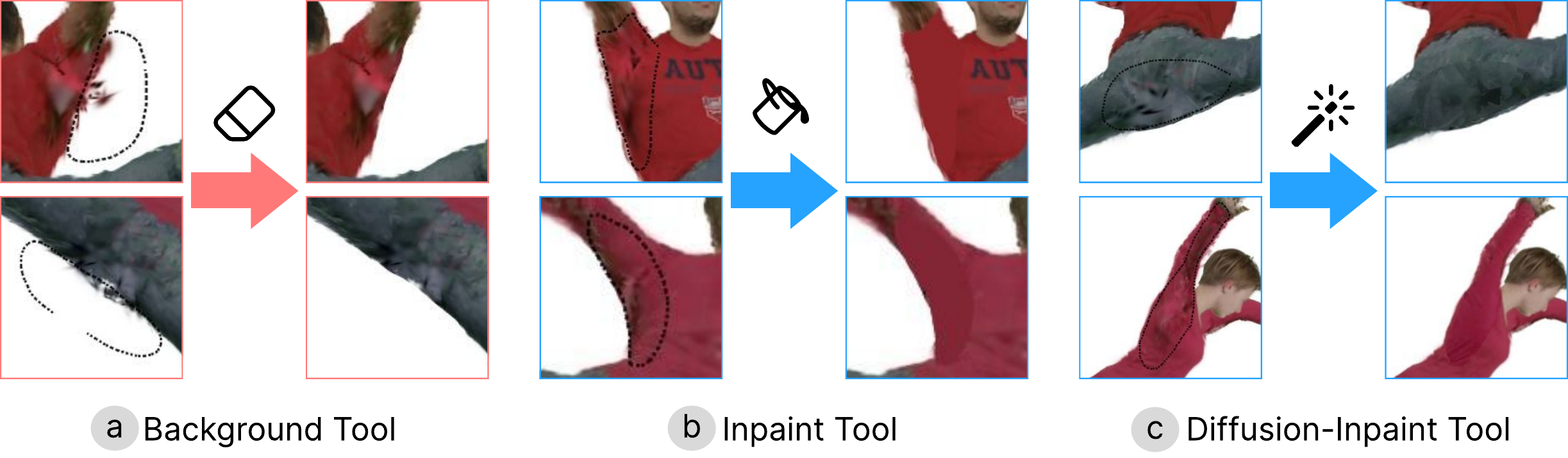}
    \caption{
    \textbf{Usage of the 2D image editing tools in \systemName.} 
    (a) The background tool allows the user to enclose areas of unwanted Gaussians in the suggested image $J_k \in \mathcal{J}$ and fill them with the background color.
    (b) With the inpaint tool, the user surrounds the area where they want to change the color on the suggested image $J_k \in \mathcal{J}$ and fills the selected area with a chosen color.
    (c) With the diffusion-inpaint tool, as with the inpaint tool, the user surrounds the area where they want to change the color on the suggested image $J_k \in \mathcal{J}$ and fills the selected area with the diffusion model.
    }
    \label{fig:user_painting}
\end{figure*}

\paragraph{Background Tool}
Users can use the background tool to paint over the regions with floating artifacts in the rendered image with the background color (\cref{fig:user_painting}a).
This tool lets users clarify the boundaries between the 3DGS avatar and the background.
Additionally, regions filled with the background tool are areas where the user has determined that no Gaussian should be present.
Therefore, \systemName internally deletes the Gaussians whose centers exist in those areas.

\paragraph{Inpaint Tool}
When using the inpaint tool, users can choose a color to correct areas with unnatural colors in the rendered image (\cref{fig:user_painting}b).
This inpaint tool allows users to effectively refine anomalous color artifacts at regions with flat textures.

\paragraph{Diffusion-Inpaint Tool} 
When artifacts appear in areas with complex textures, using the inpaint tool is not sufficient to refine the artifacts.
Therefore, we provide a 2D image inpaint tool using a pretrained diffusion model~\cite{Rombach_2022_CVPR} to synthesize suitable image content in the selected region.
As shown in~\cref{fig:user_painting}c, the user first selects a region that contains the undesired artifacts and optionally specify a text prompt that describes the desired content.
Then, the diffusion model generates three different inpainted images, allowing the user to select the most suitable one as the refined image.

\subsection{Automatic Body and Camera Pose Suggestion}
\label{sec:pose_suggestion}
Body and camera pose suggestion chooses a body and camera pose that displays as many as possible barely visible Gaussians in the original input video.
In the following paragraph, we use ``pose'' to denote both body and camera pose.
Our pose suggestion is built on the assumption that the barely visible Gaussians are more likely to generate visual artifacts in the fitted 3DGS avatar.
Therefore, our suggested pose minimizes the user's need to manually locate the problematic Gaussians and refine them efficiently.
We first define Gaussian visibility, which measures the contribution of each Gaussian to each pixel in the 2D image rendering.
We then introduce how the suggested pose is obtained based on the Gaussian visibility.

\begin{figure*}
    \centering
    \includegraphics[width=\linewidth]{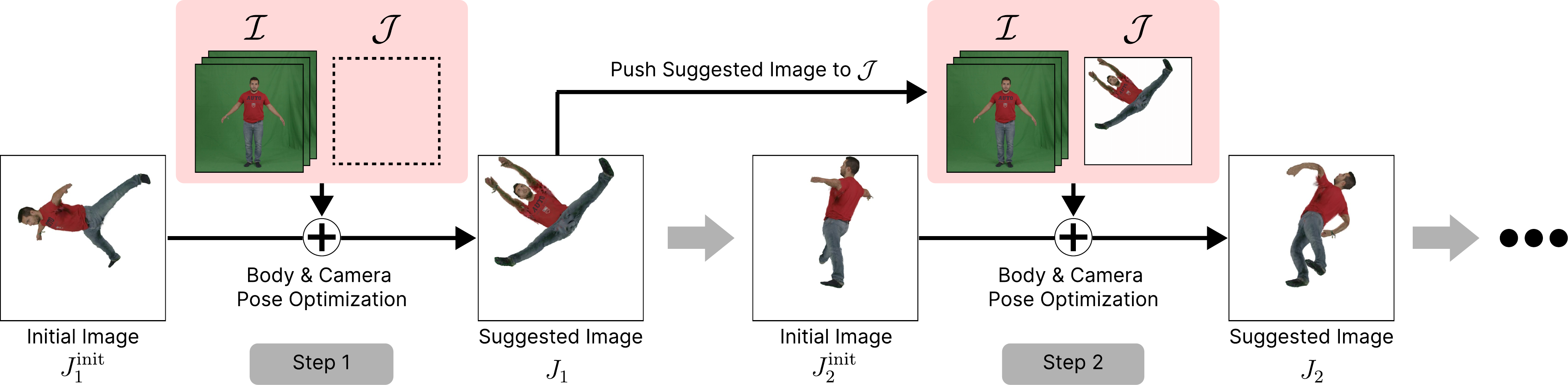}
    \caption{
    \textbf{Body and camera pose suggestion.} 
    At each step, \systemName optimizes the body and camera poses using the accumulated visibility $\mathbf{V}$ from the input video and the edited images in the previous steps.
    Artifacts are more likely to occur in the areas occluded in the input video due to the lack of information.
    Therefore, starting from the canonical body pose and a uniformly sampled camera pose, we apply the optimization to ensure that these occluded areas are included as much as possible in the suggested image $J_k$.
    This suggestion allows the user to refine the avatar with minimal image editing. 
    }
    \label{fig:pose_suggestion}
\end{figure*}

\subsubsection{Gaussian Visibility}
As described in~\cref{eq:original_color}, the color of a pixel $p$ is determined by the color of each Gaussian, the transmittance of the Gaussians in front of each one, and the value of each Gaussian.
When we regard~\cref{eq:original_color} as the weighted average of the colors of each Gaussian on the ray from the viewpoint to the pixel $p$, we can interpret the weight $T_i^p \alpha_i^p$ for each Gaussian as its contribution to pixel $p$.
This contribution represents how much of each Gaussian is visible along the ray corresponding to pixel $p$.
Therefore, we can define the visibility of Gaussian $g_i$ at a pixel $p$ as:
\begin{align}
    v_i^p = T_i^p\alpha_i^p
\end{align}

Collecting the visibility from all the pixels, we can also define the visibility of Gaussian $g_i$ on a rendered image $I = \mathcal{R}(\mathcal{G}, \theta, \tau)$ as:
\begin{align}
v_i^I = \sum_{p \in I} v_i^p 
\end{align}
We then define a function $\sigma$  to compute the visibility for all Gaussians on the rendered image $I = \mathcal{R}(\mathcal{G}, \theta, \tau)$ as:
\begin{align}
\sigma (\mathcal{G}, \theta, \tau) = [v_1^I,..., v_N^I] \in \mathbb{R}^N,
\end{align}
where $N$ represents the number of the Gaussians $\mathcal{G}$ in the original 3DGS avatar.

\subsubsection{Body and Camera Pose Suggestion Optimization}
To suggest a pose that displays as many as possible barely visible Gaussians after $K$ iterations, we start by measuring the visibilities of Gaussians in the input video $\mathcal{I} = \{(I_1, \theta_1, \tau_1), (I_2, \theta_2, \tau_2), \dots, (I_T, \theta_T, \tau_T)\}$ and the painted images $\mathcal{J} = \{(J_1, \theta_1', \tau_1'), (J_2, \theta_2', \tau_2'), \dots, (J_K, \theta_K', \tau_K')\}$ as:
\begin{align}
\mathbf{V} = (1 - w) \biggl( \frac{1}{|\mathcal{I}|} \sum_{(I, \theta, \tau) \in \mathcal{I}} \sigma(\mathcal{G}, \theta, \tau) \biggl)~+~w \biggl( \frac{1}{|\mathcal{J}|} \sum_{(J, \theta', \tau') \in \mathcal{J}} \sigma(\mathcal{G},\theta', \tau') \biggl).
\end{align}
where $J_k$ represents the $k$-th refined avatar image, while $\theta'_k$ and $\tau'_k$ denote the $k$-th suggested body and camera pose, respectively. 
We set the weight $w$ for $\mathcal{J}$ to $0.01$ in all subsequent experiments.


Starting from the canonical body pose and a uniformly sampled camera
pose, we then minimize the following function to obtain the suggested body and camera pose $(\theta_{K+1}', \tau_{K+1}') = (\mathcal{D}(z_{K+1}'), \tau_{K+1}')$ (\cref{fig:pose_suggestion}):
\begin{align}
(z_{K+1}', \tau_{K+1}') &= \argmin_{(z, \tau)}   \sum_{i=1}^{N} \text{clip}(\mathbf{V}[i] - \bar{V})\cdot \sigma(\mathcal{G}, \mathcal{D}(z), \tau)[i], \qquad \\ \notag
    \text{clip}(x) &=
    \begin{cases}
        x & \text{if } x < 0, \\
        0 & \text{if } x \ge 0
    \end{cases}, 
\end{align}
where $[i]$ represents the value of the $i$-th element in a vector, and $\mathcal{D}$ is a pose decoder~\cite{smplx} that generates a body pose from a low-dimensional latent vector $z\in\mathbb{R}^{32}$. 
The purpose of using this additional pose decoder is to prevent the optimized body pose from deviating from natural poses.
We explain the usage of the pose decoder in more detail in the supplemental material.
Intuitively, this function aims to select a body and camera pose that can display as many Gaussians as possible, which are less visible in all observed viewpoints but more visible in this one.
Therefore, users can identify artifacts without the need to explore different body and camera poses. 



\begin{figure*}
    \centering
    \includegraphics[width=\linewidth]{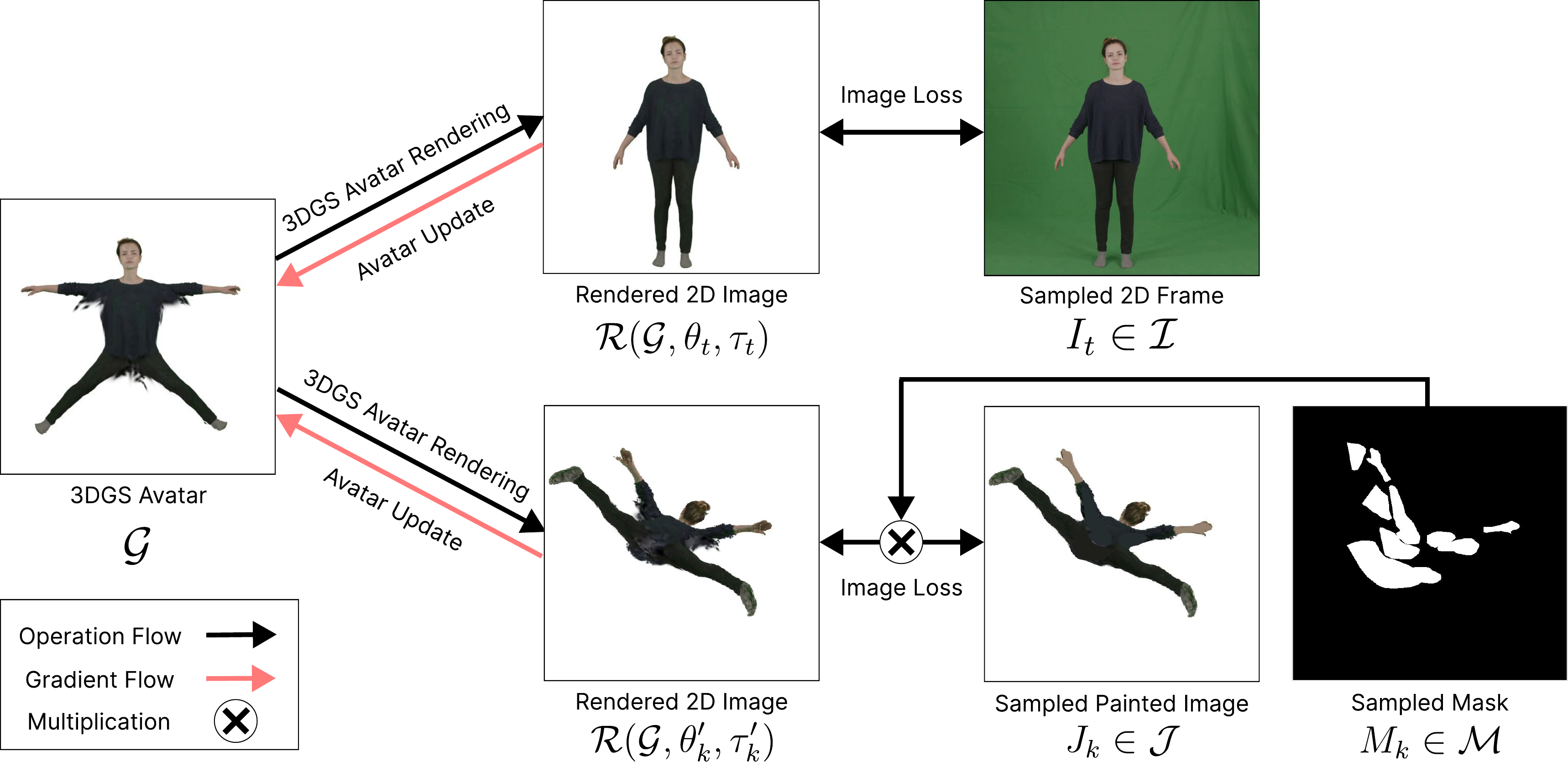}
    \caption{
    \textbf{Additional training of 3DGS avatar.}
    We conduct the additional training of the 3DGS avatar following the optimization strategy of~\cref{fig:gaussian_avatar_training}.
    In the additional training, we use both $\mathcal{I}$, the original input video, and $\mathcal{J}$, the images edited by the user.
    For each step during the additional training, we randomly determine which source to update the avatar from $\mathcal{I}$ or $\mathcal{J}$ and then compute the image loss with an image sampled from the selected source.
    When we compute the image loss between the rendered image and the sampled image from $\mathcal{J}$, we multiply the sampled binary mask $M_k$, which corresponds to the regions of the user editing.
    }
    \label{fig:gaussian_avatar_retraining}
\end{figure*}
\subsection{Additional Training of 3DGS Avatar}
\label{sec:fine-tuning}
Using the original input video $\mathcal{I}$ and the edited images: 
\begin{align}
\mathcal{J} = \{(J_1, \theta_1', \tau_1'), (J_2, \theta_2', \tau_2'), \dots, (J_{K + 1}, \theta_{K + 1}', \tau_{K + 1}')\},
\end{align}
which we add the newly edited image $J_{K + 1}$ to, we conduct the additional training for the 3DGS avatar (\cref{fig:gaussian_avatar_retraining}).
As the optimization process described in~\cref{sec:prelim}, we first randomly sample an image from $\mathcal{I}$ and $\mathcal{J}$.
We then compute the image loss $L_\text{img}$ between the rendered avatar image and the randomly sampled image.
This optimization of the 3DGS avatar for both $\mathcal{I}$ and $\mathcal{J}$ encourages the 3DGS avatar to have the same appearance as the input video $\mathcal{I}$ and the user edits $\mathcal{J}$.

To focus exclusively on the user-edited areas, we apply the binary mask $M_k$ of user editing for each suggested image. We compute the image loss $L_\text{img}$ for the sample image drawn from $\mathcal{J}$ as follows:
\begin{align}
    L_{\text{img}}(M_{k}\cdot\mathcal{R}(\mathcal{G}, \theta'_{k},\tau'_{k}), M_{k}\cdot J_{{k}}).
\end{align}
In addition, to incorporate the user's edits into the 3DGS avatar with a minimal number of iterations, we weight the image sampling from $\mathcal{J}$.
We set the probability of sampling an image from $\mathcal{J}$ during optimization to 0.3 in all subsequent experiments.

%% file: 07_result.tex
\section{Refined Avatar Results}
\begin{figure*}[h!]
    \centering
    \includegraphics[width=\linewidth]{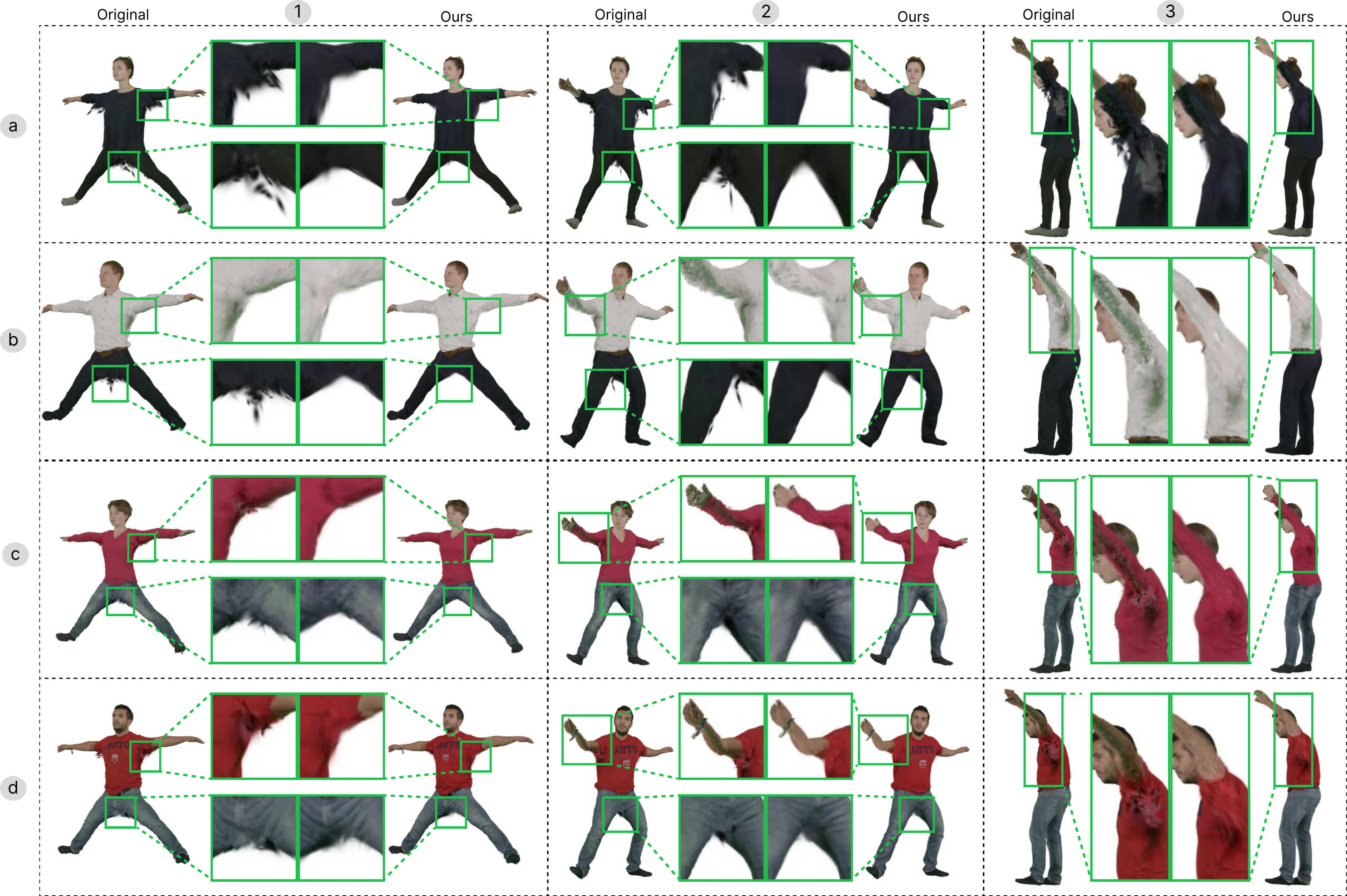}
    \caption{
    \textbf{Refined avatar examples.} 
    We show the refined avatar results in three poses.
    For each pair of avatars, the left avatar is the original avatar, and the avatar on the right is the refined avatar using \systemName.
    Users can effectively refine both floating Gaussians and anomalous color Gaussians across multiple poses.
    Each refinement required editing 5 images in less than 10 minutes.
    }
    \label{fig:refined_avatars}
\end{figure*}

In~\cref{fig:refined_avatars}, we present the refined 3DGS avatars using \systemName.
Each original avatar is generated from a monocular video in the People-Snapshot dataset~\cite{alldieck2018detailed} using a 3DGS avatar generation algorithm, GART~\cite{lei2024gart}.
Afterward, the same user refined each 3DGS avatar by editing five images rendered with the body and camera poses suggested by \systemName.
The entire process, including body and camera pose suggestions and additional avatar training, took less than $10$ minutes per avatar.

As shown in~\cref{fig:refined_avatars}, the original avatars have floating Gaussians around the armpit and crotch areas.
The user successfully used \systemName to remove them effectively and obtain visual appealing rendered results.
Moreover, with \systemName, The user successfully refined the anomalous color Gaussians around hidden areas, such as the inner part of the arm.
These results suggest that users can use the 2D image editing tools provided by \systemName to effectively refine both the floating and the anomalous color Gaussians in a short period of time.

%% file: 08_study.tex
\section{User Study}
We conducted a user study to understand the effectiveness of \systemName in refining 3DGS avatars quantitatively and qualitatively.
To better understand the benefits of \systemName, we prepared a baseline interface for comparison: 3DGS editor \textit{SuperSplat}~\cite{supersplat}.
The participants were asked to refine 3DGS avatars using both systems and provide feedback through questionnaires about their experience.



\subsection{Study Setup}
\subsubsection{Participants.}
We recruited four participants (P1 - P4, four male) aged 21 - 26 from the computer science department.
All the participants had intermediate knowledge of the field of user interface, and three of them had intermediate knowledge of the field of computer graphics.
Additionally, except for one participant, the participants had no experience in building and using 3DGS avatars. 
They received an Amazon Gift Card worth $1,500$ JPY ($< 15$ USD) as compensation for participating in the user study.


\subsubsection{Baseline 3DGS Avatar.}
We used GART~\cite{lei2024gart} to generate the initial 3DGS avatars from monocular videos.
We set the number of training iterations to $2,000$ for all avatars used in our user study.
For the additional training of the 3DGS avatars after each user refinement iteration in our system, we set the number of training iterations to $500$.
Overall, we used two 3DGS avatars, one male avatar, and one female avatar, in our user study, and each avatar was refined by four different participants.

\subsubsection{Baseline Interface.}
We compared \systemName with a baseline system: SuperSplat~\cite{supersplat}.
SuperSplat allows users to rotate or translate all Gaussians simultaneously and delete the selected Gaussians. 
In our user study, we only provided the Gaussian deletion feature because the simple transformation of all Gaussians cannot contribute to refining the avatars.
When using \textit{SuperSplat}, users first identify and select the Gaussians that cause artifacts.
It's important to note that \textit{SuperSplat} does not allow users to change the avatar's body pose. 
Therefore, we kept the body pose fixed in the canonical pose as shown in~\cref{fig:gaussian_avatar_training} and the user edited the 3DGS avatars by adjusting only the camera pose.

In \textit{SuperSplat}, there are two erasing tools that allow users to delete unwanted Gaussians to refine the 3DGS avatars.
For both erasing tools, the users first select an area that contains unwanted Gaussians.
We implemented this user interface for selecting regions in the same way as the user interface for 2D image editing in \systemName.
When users choose to use the ``Erase Center Tool'', they can delete Gaussians whose centers are inside the selected regions.
On the other hand, when users choose to use the ``Erase Splat Tool'', they can remove Gaussians that overlap with the selected regions. 
In the implementation, Gaussians with Gaussian visibility greater than 0.01 at a given pixel in the selected regions were considered overlapping.

\subsubsection{Dataset.}
\label{sec:study_setup_dataset}
To assess the quality of the 3DGS avatars refined by the users, it is essential to have a reference video for each refined avatar. 
For this purpose, we utilized the iPER dataset~\cite{lwb2019}, which provides two videos per actor: one where the actor rotates in an A-pose and another where the actor performs free motions.
We used the A-pose video as training data to generate the 3DGS avatars with artifacts and the free motion video as the ground truth for evaluating the refined avatars.
To estimate the actor's pose and camera pose in each frame, we applied OpenPose~\cite{lin2023one} during the training and evaluation phases.

For our user study, we selected two actors from the iPER dataset: Avatar 1 (001\_2, male) and Avatar 2 (007\_3, female).
We excluded clothing with text or complex patterns because neither SuperSplat~\cite{supersplat} nor \systemName can effectively handle or refine them.
Additionally, we excluded loose-fitting clothing, open jackets, and videos shot outdoors, as avatar quality from such videos was insufficient.
To generate the 3DGS avatars, we sampled $100$ frames every three frames from the beginning of the video.


\subsubsection{Hardware Setup}
All the participants conducted our user study using a desktop PC with an Intel(R) Core(TM) i9-10900F CPU and NVIDIA GeForce RTX 3080 GPU that can effectively train and render 3DGS avatars.


\subsection{Procedure}
The user study procedure was as follows. 
First, we gave participants an overview of the experiment.
Next, they participated in two sessions: one using \textit{SuperSplat} and the other using \systemName. 
We started by giving participants instructions on how to use the first assigned system.
They then practiced using this system for five minutes.
Next, we asked the participants to refine the first avatar with the first assigned system.
Once the participants finished refining the first avatar, we asked them to complete a questionnaire regarding their experience using the first assigned system.
After that, we provided instructions on the second system and allowed them to practice using it.
We then asked the participants to refine the second avatar using the second assigned system.
Finally, we asked the participants to complete a questionnaire on their experience using the second assigned system.
In total, each participant refined two avatars using different systems and it took about $45$ minutes to finish the experiment.

Following the within-subjects design, we used a different order of system and avatar assignment between each participant to counterbalance the order of exposure.
There are four possible orders in which participants conduct the experiments: the order in which they use the \textit{SuperSplat} and \systemName, and the order in which they refine Avatar 1 and Avatar 2.
Therefore, we assigned the four order patterns to the four participants as illustrated in~\cref{tab:assignment_order}. 
Note that we used the same avatar for two practice sessions, which we did not use in the experiment.

\begin{table}[]
    \centering
    \begin{tabular}{l|c|c|}
        & Session 1 & Session 2 \\
    \hline
    P1   & \textit{SuperSplat}~-~Avatar 1 & \systemName~-Avatar 2 \\
    P2   & \textit{SuperSplat}~-~Avatar 2 & \systemName~-~Avatar 1 \\
    P3   & \systemName~-~Avatar 1 & \textit{SuperSplat}~-~Avatar 2 \\
    P4   & \systemName~-~Avatar 2 & \textit{SuperSplat}~-~Avatar 1\\
    \hline
    \end{tabular}
    \caption{
    \textbf{Order of method and avatar assignment of user study.}
    We assigned the order of method and avatar assignment to counterbalance the exposure order for the methods and avatars. 
    }
    \label{tab:assignment_order}
\end{table}

During the editing process, we provided participants with the assigned system and the corresponding input video used to generate the 3DGS avatar. 
Participants examined the video and refined the artifacts on the 3DGS avatars to make them resemble the actor in the video.
Users were allowed to perform five iterations of refinement.
In each iteration, when using \textit{SuperSplat}, users could select one desired viewpoint and delete any number of Gaussians they wanted to erase.
On the other hand, when using \systemName, users could refine the presented avatar image using any painting tools mentioned in~\cref{sec:system_ui} in each iteration.
We set a maximum refinement time of $10$ minutes for both systems, and we terminated the session if the time exceeded $10$ minutes. 
When using \systemName, if the session was terminated before users can finish all five iterations, we still performed the remaining additional training of the avatars.
The purpose of performing these remaining training is to ensure that all avatars refined by \systemName have the same total training iteration numbers.

After each avatar refinement session, the participants were asked to complete a questionnaire regarding their experience using either \textit{SuperSplat} or \systemName, including its usability and perceived usefulness. 
In the questionnaire, we started by asking usability questions related to this task, followed by questions about general tool usability according to SUS~\cite{brooke1986system}.
All questions in the questionnaire were on the Likert $5$-point scale from ``Strongly disagree'' to ``Strongly agree'' (from $1$ to $5$) or from ``Not useful at all'' to ``Extremely useful'' (from $1$ to $5$).
We also measured the total time taken for refining avatars using both the \textit{SuperSplat} and \systemName, as well as the time required for the additional training of 3DGS avatars.



\subsection{Results}

\subsubsection{Quality Evaluation.}
We assessed the quality of the 3DGS avatars updated during the user study by introducing metrics for floating Gaussians and anomalous color Gaussians. 
In this quantitative evaluation, we compared the refinement results using \systemName and \textit{SuperSplat}~\cite{supersplat} by using the corresponding free motion videos from the iPER dataset.
Specifically, we sampled frames from the free motion videos in the same manner we sampled the initial training images from the A-pose video, as described in~\cref{sec:study_setup_dataset}.

\begin{figure*}[h!]
    \centering
    \includegraphics[width=\textwidth]{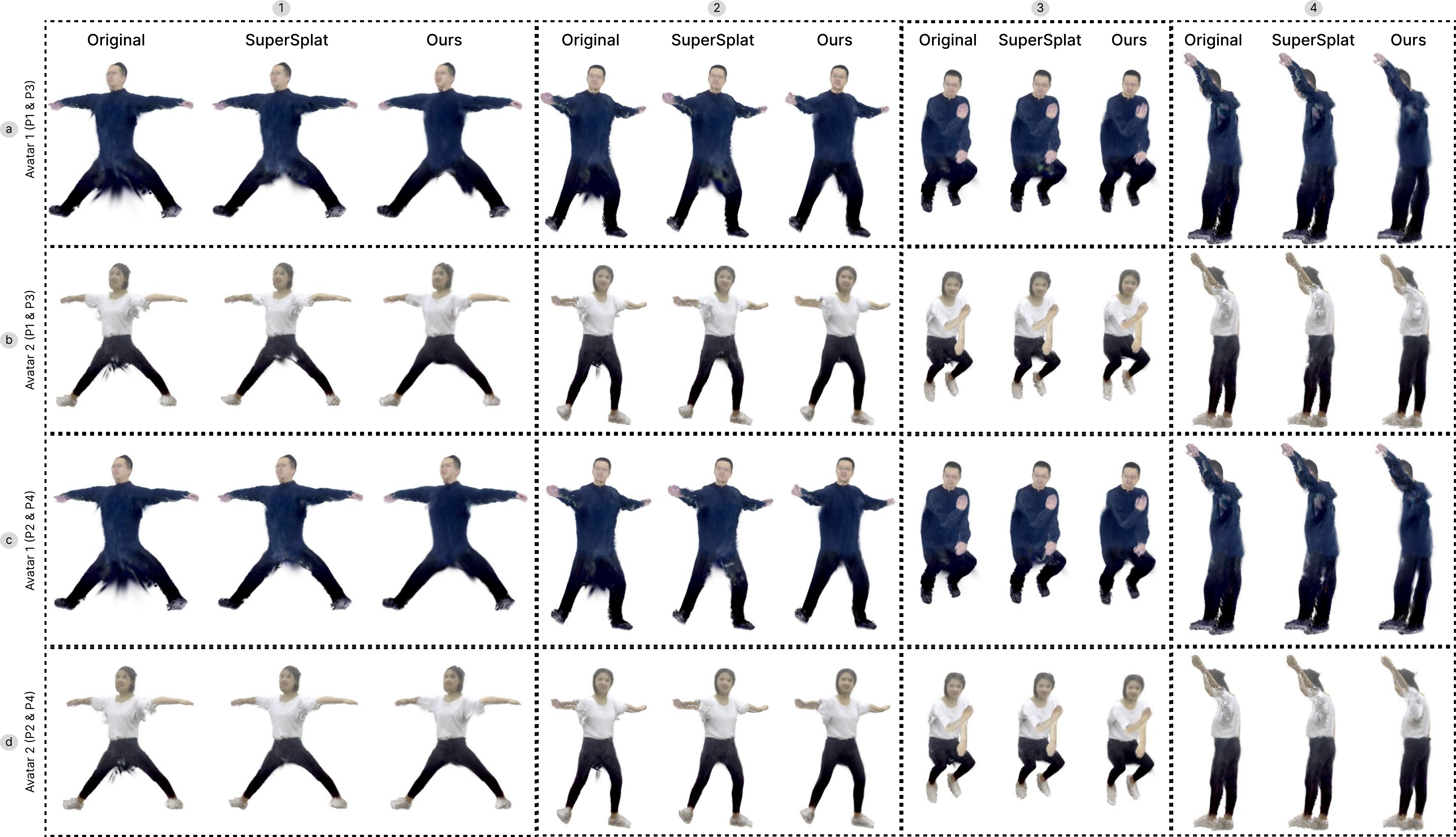}
    \caption{
    \textbf{Avatars edited in user study.}
    We show the original avatars and refined avatars by the user study participants.
    For each pair of avatars, the avatar on the left is the original, unedited version, the center avatar was edited using SuperSplat~\cite{supersplat} during the user study, and the avatar on the right was edited using \systemName.
    }
    \label{fig:user_edit_avatar}
\end{figure*}

We employed IoU (Intersection-over-Union)~\cite{yu2016unitbox} for evaluating the quality of floating Gaussian refinement.
First, we applied Segment Anything Model (SAM)~\cite{kirillov2023segany} to obtain the binary mask of the actor of each frame in the free motion video.
Next, we calculated the Gaussian visibility for all pixels of the rendered images of the refined 3DGS avatars, and we only considered pixels with the $\sum_{i \in \mathcal{N}_p} v_i^p$, the accumulated visibility more than $0.5$ as part of the mask of the 3DGS avatar.
Finally, we calculated the IoU between the actor mask from the free motion video and the 3DGS avatar to assess whether the floating Gaussians had been appropriately refined.
In~\cref{tab:user_quan}, for each avatar, we showed the mean IoU of two refined results.
It shows that the avatars refined by our method achieve higher IoU compared to avatars refined by \textit{SuperSplat}.
Moreover, we can see that users could smoothly interpolate the regions surrounding the deleted Floating Gaussians using \systemName (\cref{fig:user_edit_avatar}). 
Additionally, when Gaussians are directly deleted with the \textit{SuperSplat}, previously hidden anomalous color Gaussians may become visible (\eg~\cref{fig:user_edit_avatar}a2, c2).
However, since \systemName performs the additional training after each user refinement, \systemName is less likely to exhibit such Gaussians (\cref{fig:user_edit_avatar}).

\begin{table*}[htbp]
    \centering
    \begin{tabular}{l|ccc|ccc|}
    & \multicolumn{3}{c|}{Avatar 1 (male)} & \multicolumn{3}{c|}{Avatar 2 (female)} \\
    \cline{2-7}
    & IoU$\uparrow$ & PSNR$\uparrow$ & SSIM$\uparrow$ & IoU$\uparrow$ & PSNR$\uparrow$ & SSIM$\uparrow$ \\
    \hline

    SuperSplat~\cite{supersplat} & 0.9094 & 23.058 & 0.9403 & 0.8636 & 25.604 & 0.9555 \\ 
    Ours & \bestcell{0.9115} & \bestcell{23.463} & \bestcell{0.9479} & \bestcell{0.8713} & \bestcell{26.014} & \bestcell{0.9593} \\ 

    \hline
    \end{tabular}
    \captionsetup{width=\textwidth}
    \caption{\textbf{Quantitative results on refined avatars.} 
    We show the mean metrics for both avatars used in our user study. The refined avatars using our method obtained better results compared to SuperSplat~\cite{supersplat} for both avatars.
    We highlight the \besthint{best} result.
    }
    \label{tab:user_quan}
\end{table*}

On the other hand, we employed image similarity metrics, SSIM and PSNR, to evaluate the refinement of the anomalous color Gaussians. 
First, we used SAM~\cite{kirillov2023segany} to obtain a background mask for each frame in the free motion video. 
Using the extracted background mask, we created a ground truth image for each frame by replacing the original background with a white one.
Then, we calculated the image similarity metrics between the ground truth images and the rendered avatar images.
Similar to the IoU metric, for each avatar, we showed the mean SSIM and PSNR of two refined results in~\cref{tab:user_quan}.
It shows that the avatars refined by our method obtain better image similarity metrics (\cref{tab:user_quan}).
As shown in~\cref{fig:user_edit_avatar}, we can also observe that the anomalous color Gaussians are better removed by \systemName (\eg~\cref{fig:user_edit_avatar}b2, c4).

\subsubsection{Manual Operation Time}

We categorized the refinement process of both systems into three parts: \textit{Pose Decision}, \textit{2D Editing}, and \textit{Avatar Update}. 
We presented the time spent on each of the these categories, as well as the total time taken to edit each avatar in~\cref{tab:manipulation_time}.

The \textit{Pose Decision} category refers to the time taken to determine the body and camera pose. 
When using \textit{SuperSplat}, users manually selected only the camera pose, while \systemName automatically suggested both the body and camera poses through optimization techniques.
As shown in~\cref{tab:manipulation_time}, users spent less time determining a pose to refine when using \systemName compared to manually selecting a pose using \textit{SuperSplat}.
The \textit{2D Editing} category represents the time spent by users editing the avatars on the 2D screen.
Overall, participants spent more time performing 2D editing with \systemName.
This is because \systemName offered the ability to fill in the anomalous color Gaussians, while \textit{SuperSplat} only allowed users to remove Gaussians.
This broad range of editable content led to an increase in the number of edits, resulting in longer 2D editing times.
Lastly, the \textit{Avatar Update} category refers to the time \systemName required to perform additional avatar training with the refined 2D images. 
In the user study, each avatar underwent five additional trainings, with an average of $26$ seconds per session.
In total, this amounted to approximately $130$ seconds for all additional training, as shown in~\cref{tab:manipulation_time}.

Although users took longer to refine the avatars using \systemName, this was mainly due to the broader range of operations available in 2D image editing provided by \systemName and the additional training required after each 2D image editing session.
If we consider only the interaction time (\ie~exclude the additional training time), users only need to spend a small amount of extra time, and they can obtain avatars in higher quality (\cref{tab:manipulation_time}e).
Meanwhile, we believe that our system will benefit from future developments in 3DGS rendering speeds to reduce additional learning time.

\begin{table*}[]
\centering

\begin{tabular}{l|l|c|c|c||c|c|}
\multicolumn{2}{c|}{} & (a) Pose Decision & (b) 2D Editing & (c) Avatar Update & (d) Total & (e) Interaction\\ 
\hline 
\multirow{2}{*}{Avatar 1} & \textit{SuperSplat}~\cite{supersplat} & 103~$\pm$~9.19 & 277~$\pm$~74.25 & NaN & 380~$\pm$~83.44 & 380~$\pm$~83.44 \\ 
 & Ours & 35~$\pm$~2.83 & 494~$\pm$~48.08 & 129~$\pm$~10.61 & 658~$\pm$~61.52 & 494~$\pm$~48.08 \\ 
\hline 
\multirow{2}{*}{Avatar 2} & \textit{SuperSplat}~\cite{supersplat} & 120~$\pm$~48.08 & 319~$\pm$~149.20 & NaN & 439~$\pm$~197.28 & 439~$\pm$~197.28 \\ 
 & Ours & 34~$\pm$~1.41 & 471~$\pm$~7.78 & 130~$\pm$~4.95  & 635~$\pm$~14.14 & 471~$\pm$~7.78 \\ 
\hline 
\end{tabular}
\caption{
    \textbf{Manual Operation time.}
    The numbers in each cell represent the mean and standard deviation of the total time spent in each category.
    (a) The \textit{Pose Decision} value represents the total time taken to determine the body and camera poses.
    (b) The \textit{2D Editing} value represents the total time the user spent editing on the 2D screen.
    (c) Finally, the \textit{Avatar Update} column indicates the total time required for the additional avatar training after each 2D image edit in \systemName.
    The \textit{Total} column in (d) represents the sum of the values in columns (a), (b), and (c), while the \textit{Interaction} column in (e) indicates the time the user spent interacting with each system.
    For interaction time, we recorded the sum of columns (a) and (b) for \textit{SuperSplat}, and column (b) for our system.
    All values are expressed in seconds.
}
\label{tab:manipulation_time}
\end{table*}

\subsubsection{Usability Rating}
The average SUS score for \systemName was $80.63~\pm~17.87$.
According to the response to SUS-based questions, all participants agreed that the \systemName was not unnecessarily complex and that using \systemName was not cumbersome.
Additionally, all participants answered that the functionality of \systemName was well integrated, and they felt most people would be able to use the system quickly.

\begin{figure}[!htb]    
    \begin{minipage}[c]{.48\textwidth}
        \centering
        \includegraphics[width=\linewidth]{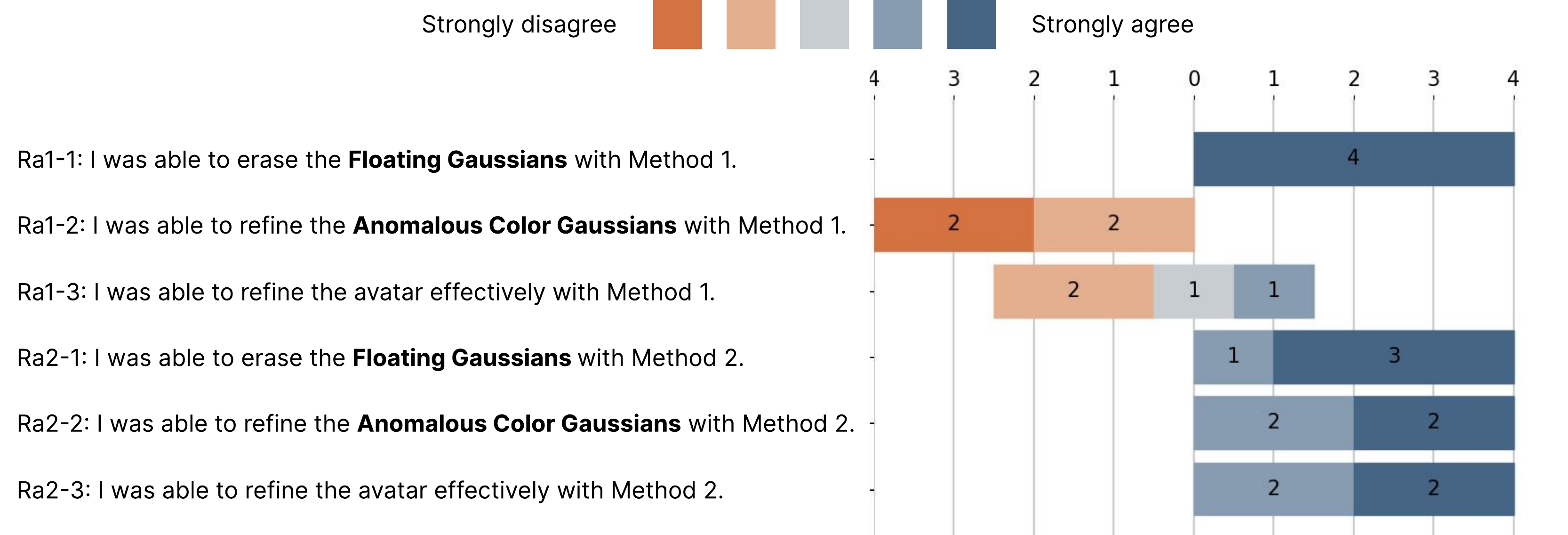}
    \end{minipage} \quad
    \begin{minipage}[c]{0.48\textwidth}
        \centering
        \includegraphics[width=\linewidth]{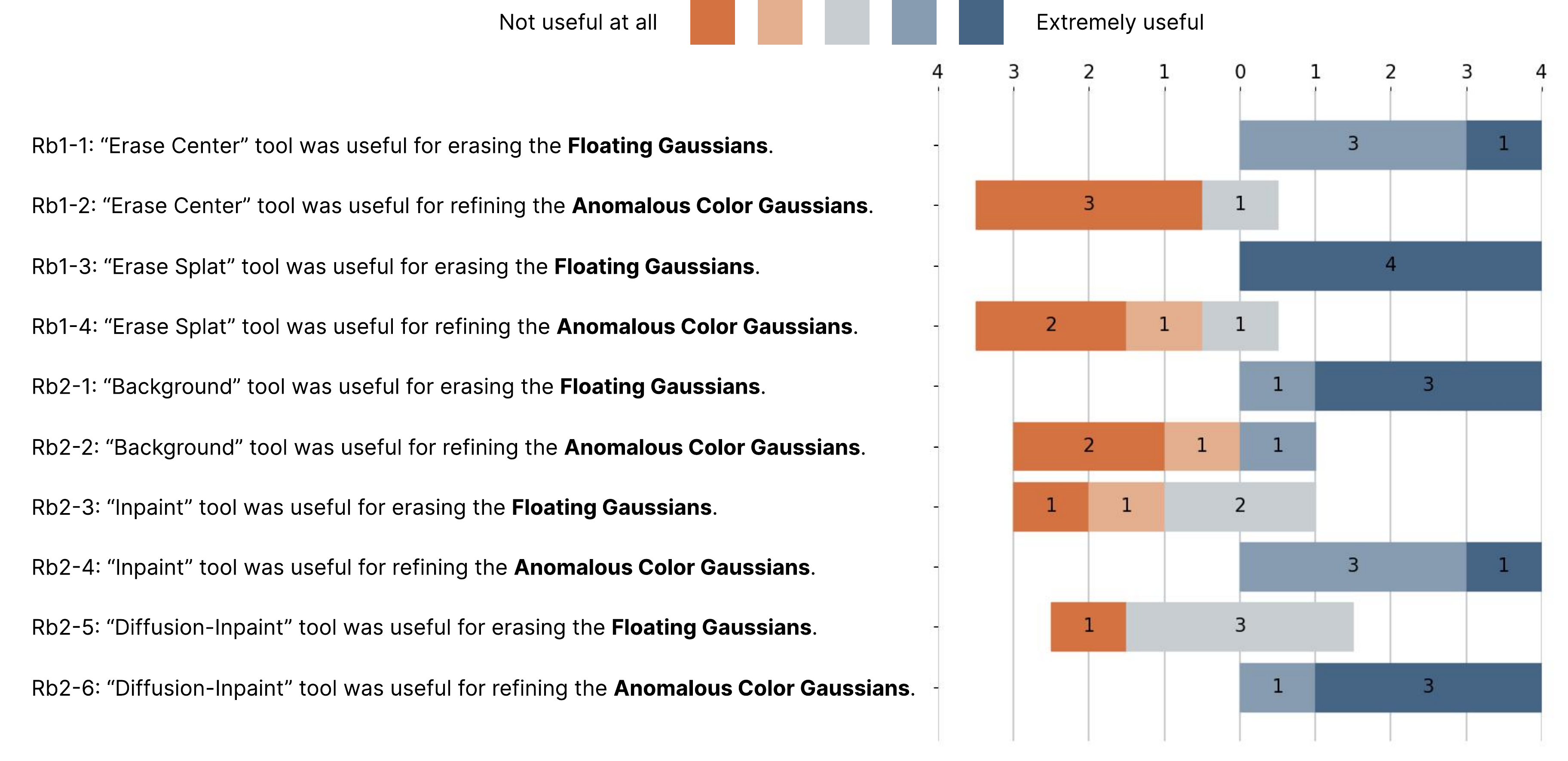}
    \end{minipage}
    \begin{minipage}{.48\textwidth}
      \subcaption{\textbf{Usability rating on whole systems.} For each question, we asked the participants to rate the systems from "Strongly disagree" to "Strongly agree", following the Likert 5-point scale. In addition, Method~1 represents the baseline system, \textit{SuperSplat}, while Method~2 represents our system, \systemName.}\label{fig:ra_result}
    \end{minipage} \quad
    \begin{minipage}{.48\textwidth}
      \subcaption{\textbf{Usability rating on tools.} For each question, we asked the participants to rate each tool from "Not useful at all" to "Extremely useful", following the Likert 5-point scale.
      Participants answered each question from the perspective of the Floating Gaussians and the Anomalous Color Gaussians respectively.}\label{fig:rb_result}%
    \end{minipage}
    \caption{\textbf{Usability ratings.}}%
    \label{fig:imgs}%
\end{figure}

First, we presented the results of the system usability questionnaire in~\cref{fig:ra_result}.
In~\cref{fig:ra_result}, ``Method 1'' refers to \textit{SuperSplat} and ``Method 2'' refers to \systemName.
Overall, \textit{SuperSplat} received higher ratings for its effectiveness in removing the floating Gaussians, but lower ratings for addressing the anomalous color Gaussians.
This is because \textit{SuperSplat} directly removes Gaussians, making it challenging to edit color-related artifacts like the anomalous color Gaussians.
In contrast, all users found \systemName effective in handling both the floating Gaussians and anomalous color Gaussians.

Next, we showed the questionnaire results on the usability for each tool provided by \textit{SuperSplat} and \systemName in~\cref{fig:rb_result}.
For removing the floating Gaussians, the erase center tool and the erase splat tool from \textit{SuperSplat} received high ratings, as well as the background tool from \systemName.
Among these tools, the erase splat tool received the highest rating.
However, the inpaint tool and the diffusion-inpaint tool provided by \systemName received lower ratings for removing floating Gaussians.
Regarding refining anomalous color Gaussians, the participants rated highly the inpaint tool and the diffusion-inpaint tool from \systemName, with the diffusion-inpaint Tool receiving the highest rating.
In contrast, both tools provided by~\textit{SuperSplat} and the background tool received lower ratings for addressing anomalous color Gaussians.

\subsection{Crowdsourced Quality Evaluation}
To further assess the quality of the avatars refined by the user study participants, we recruited $30$ unpaid volunteer evaluators ($17$ male, $13$ female) between the ages of 21-51 for an online crowdsourcing study.
This study aimed to have evaluators rate the same avatar refined by two different systems.
According to~\cref{tab:assignment_order}, we can obtain two sets of refined avatars for each avatar, resulting in four pairs of avatars.
Each pair of avatars was then animated with three motions, including fencing, eating, and dancing from GART~\cite{lei2024gart}.
This resulted in $12$ unique avatar pairs.
In each question, we present the animation of one avatar pair to the crowdsourced evaluators and asked them to select the avatar they believed had better quality and fewer artifacts throughout the animation.
Overall, we had $12$ questions.

To reduce the burden on evaluators, we divided the $12$ questions into two groups: set 1 and set 2, each contains six questions.
To prevent random answers from potentially malicious evaluators, we included two identical questions at the beginning and end of the questionnaire with swapped order.
We discarded all answers from evaluators who answered both of these two questions inconsistently.
We randomly assigned the evaluators to each question set.
For set 1, $16$ evaluators completed the questions, with two evaluators filtered out based on the criteria.
For set 2, $14$ evaluators completed the questions, and no participant was excluded.
Please check the question used in our crowdsourcing study and all avatar pairs in the supplemental material.

As shown in~\cref{tab:study_quan_result}, avatars edited with \systemName received higher quality ratings in $11$ out of $12$ questions.
The only case where the avatar edited with \textit{SuperSplat} received a higher quality rating was for Avatar 1 with fencing motion (P1 \& P3) (\cref{fig:user_edit_avatar}a2).
This was because there were still some artifacts present in the left leg of the avatar edited with \systemName.

\begin{table*}[htbp]
\centering
\footnotesize
\begin{tabular}{l|ccc|ccc|ccc|ccc|}
 & \multicolumn{6}{c|}{Question Set 1 (n = 14)} & \multicolumn{6}{c|}{Question Set 2 (n = 14)} \\ \cline{2-13}
 & \multicolumn{3}{c|}{Avatar 1 (P1 \& P3)} & \multicolumn{3}{c|}{Avatar 2 (P1 \& P3)} & \multicolumn{3}{c|}{Avatar 1 (P2 \& P4)} & \multicolumn{3}{c|}{Avatar 2 (P2 \& P4)}  \\ \cline{2-13}
& fencing & eating & dancing & fencing & eating & dancing & fencing & eating & dancing & fencing & eating & dancing  \\ \hline
SuperSplat~\cite{supersplat} & \bestcell{71.4} & 7.1  & 7.1  & 0.0   & 0.0   & 0.0   & 21.4 & 0.0  & 21.4 & 0.0   & 7.1  & 7.1  \\
Ours & 21.4 & \bestcell{92.9} & \bestcell{57.1} & \bestcell{100.0} & \bestcell{100.0} & \bestcell{100.0} & \bestcell{78.6} & \bestcell{85.7} & \bestcell{78.6} & \bestcell{100.0} & \bestcell{85.7} & \bestcell{92.9} \\ \hline
Same Quaity & 7.1  & 0.0  & 35.7 & 0.0   & 0.0   & 0.0   & 0.0  & 14.3 & 0.0  & 0.0   & 7.1  & 0.0 \\
\hline
\end{tabular}
\caption{
\textbf{Qualitative results on refined avatars.}
We conducted a qualitative comparison of the systems by asking evaluators to select a higher quality avatar from an avatar refined with SuperSplat~\cite{supersplat} and an avatar refined with~\systemName.
The avatars used in this study were all refined in the user study.
The number of each cell represents the percentage of the evaluators who selected the avatar edited with the corresponding system.
The number of each cell in the row of "Same quality" also represents the percentage of the evaluators who answered the quality of avatars is the same.
We highlight the \besthint{best} result.
}
\label{tab:study_quan_result}
\end{table*}

%% file: 09_discussion.tex
\section{Limitations and Future Work}

In this section, we discuss the limitations of our approach and potential future directions.

\paragraph{Multi-View Consistency of User Input}

Inconsistencies between the original input video and the edited image may prevent the user's edits from being accurately reflected on the 3DGS avatar.
This issue is especially problematic when editing shadows embedded in the avatar itself.
We treat these shadows as a type of the anomalous color Gaussian, as they create an unnatural appearance when we animate the avatar in novel poses.
However, since the original input video captures these embedded shadows, the 3DGS avatar learns the shadows as part of the Gaussian colors.
Therefore, even if the user corrects the shadows through 2D image editing, it can be difficult to properly refine them due to inconsistency with the original input video.
In the future, we plan to implement a function that detects inconsistencies between the original input video and user edits, prompting the user to determine whether the input video or the user edit should be used for the additional training.

\paragraph{Guidance for User Editing}
Inconsistencies in avatar silhouettes across user-edited images can introduce new Floating Gaussians.
This situation is particularly likely when editing avatars wearing loose clothing. 
If the actor in the original input video is wearing loose-fitting clothing, the shape of the avatar's clothing generated from the video will change significantly with changes in body pose.
\systemName, which proposes a wide range of body poses, struggles to maintain the consistency of clothing shapes between the edited images when handling avatars with such loose clothing.
Consequently, the lack of consistency in the clothing shape across the edited images can lead to the introduction of new artifacts.
To mitigate these inconsistencies, in the future, we plan to display the range of silhouette variations consistent with the user's input as a form of guidance on each suggested image.


\paragraph{Specificity to Gaussian Splatting}
Currently, our approach is tailored to 3DGS avatars. 
Extending our method to other 3D representations like mesh, NeRF, and Signed Distance Field (SDF) is a challenging task because our method depends on the Gaussian visibility to suggest body and camera poses.
In the future, we plan to investigate how to assess the visibility of various 3D representations and extend our method to refine avatars using these representations.

\paragraph{Support Direct 3D Editing through 2D Stroke}
The 2D image editing tools in our system enable users to refine the avatars easily, but they currently lack the ability to provide direct 3D guidance.
As a result, users cannot directly add more 3D Gaussians to the areas where they are needed.
In the future, we plan to incorporate a depth-aware 2D stroke that will enable users to add more Gaussians to address anomalous color Gaussian artifacts more effectively.

\paragraph{Support More 2D Image Editing Tools}

Another challenge with \systemName is the difficulty in applying edits to text or complex patterns on clothing.
Currently, \systemName provides only three editing tools for users, the inpaint tool and the diffusion-inpaint tool for editing Gaussian colors.
However, these simple tools struggle to accurately modify text and intricate patterns on clothing.
In the future, we aim to explore solutions by developing tools capable of handling more complex image edits.
Additionally, we plan to investigate the effectiveness of the diffusion-inpaint Tool, which currently relies on a pre-trained Diffusion Model, when fine-tuned specifically on the input video.


%% file: 10_conclusion.tex
\section{Conclusion}
We presented \systemName, a novel system that helps users refine 3DGS avatars generated from monocular video, addressing common artifacts caused by insufficient input data. 
Our system enables users to refine artifacts on 3DGS avatars through 2D image editing, rather than directly editing Gaussians in 3D space.
Additionally, our system suggests body and camera poses which help users minimize the number of image edits required, by leveraging the characteristics of 3DGS.
The results of our user study demonstrated that \systemName significantly enhances the visual quality of the resulting 3DGS avatars.
This work paves the way for more accurate and user-friendly 3DGS avatar creation techniques, which we hope highlights the value of integrating user input with automated processes.


